\documentclass[12pt]{article} \usepackage{subfigure} \usepackage{cite}
\usepackage{latexsym} \usepackage{amssymb} \textwidth 16cm
\begin{document}
\title{Relaxation properties of small-world networks}
\author{S. Jespersen$^{a,b}$, I. M. Sokolov$^{b}$ and A. Blumen$^{b}$ \\$^a$Institute of Physics and Astronomy \\
  University of 
{\AA}rhus, DK-8000 {\AA}rhus C, Denmark\\$^b$Theoretische Polymerphysik\\
Universit\"at Freiburg, D-79104 Freiburg i.Br., Germany}

\date{\today}  \maketitle

\begin{abstract}
Recently, Watts and Strogatz introduced the so-called {\it
small-world networks} in order to describe systems which combine
simultaneously properties of regular and of random lattices. In this
work we study diffusion processes defined on such structures by
considering explicitly the probability for a random walker to be
present at the origin. The results are intermediate between the
corresponding ones for fractals and for Cayley trees.  
\end{abstract}
\parbox{1\linewidth}{PACS numbers: 05.40.Fb,  05.60.-k, 71.55.Jv}

\section{Introduction}
Networks of the real world often seem to combine aspects from regular
and from completely random lattices. Social networks,  neural
networks, electrical powergrids, and traffic networks
\cite{watts,amaral,milgram} are all examples of patterns not described
satisfactorly by conventional regular  lattices, nor by completely
random lattices. Social structures, for instance, do not behave as
regular lattices, since (as is well known) randomly chosen people are
connected in general by a small number of intermediary bilateral
ties. Here, as in random graphs, the minimal (chemical) distance
between any two points in the system scales logarithmically with the
system size \cite{bollobas}.

To combine these two properties, Watts and Strogatz recently
introduced the idea of {\it small-world networks} \cite{watts}. This
construction is a superposition of a regular lattice with a random
lattice, and includes simultaneously well defined local clusters and
short global connections. As we will demonstrate, these systems also
display properties intermediate between those of regular and tree-like
(loop-less) lattices, already under a small number of global
connections, provided the system size is large enough.

Much work has already been done on the properties of small-world
networks
\cite{watts,amaral,barrat,monasson,newman1,moukarzel2,lago,newman2,moore,moukarzel1,barthelemy,pandit}
but most of it has focused on static (geometric) properties.  We shall
not address these issues, but rather concentrate on a {\em dynamical}
model defined on the structure. Treatments of the dynamics of
small-world networks include for instance the study of an Ising model
defined on the lattice  \cite{barrat}, spectral properties of the
small-world Laplacian, \cite{monasson}, percolation \cite{newman1},
spreading of diseases \cite{moukarzel2} and neural networks
\cite{lago}. In the following we will examine the properties of random
walks on small-world networks, in particular the relaxation,
exemplified by the probability for a random walker of being at the
original site at a later time. This is a simple quantity to extract
numerically, and very relevant for various physical properties: It is
sensitive to the topology of the network, and is related to its
vibrational modes.

\section{Definition of the model and presentation of the results.}
The small-world networks we consider are built as follows: We start
from a regular lattice with $L$ vertices in $1$ dimension under
periodic boundary conditions, each site being connected symmetrically
to its $2k$ nearest neighbours,  i.e. having as coordination number
$z=2k$. Then we add to each of the sites with probability $p$ a new
bond. The other end gets attached with equal probability to any of the
lattice sites; this allows also the possibility of vertices to become
connected to themselves. In this way we add, independent of $k$, on
the average $pL$ new bonds to the underlying regular lattice.

This construction follows \cite{newman1} for $k=1$ and is simpler than
the original procedure \cite{watts}, by which one rewires with
probability $p$ each of the original $kL$ bonds randomly.

A step-wise diffusion process is now defined by specifying all the
transition probabilities $W_{i,j}$ entering the master-equation:
\begin{equation}
\label{master}
P(i,n+1)-P(i,n)=\sum_j W_{i,j}P(j,n)-P(i,n)\sum_j W_{j,i}
\end{equation}

\noindent The $W_{f,i}$ is the probability to go from site $i$ to site
$f$ during one time step, and the probability $P(i,n)$, $i=1\ldots L$
is just the probability of being at site $i$ after the $n$th step. The
process defined in Eq. (\ref{master}) is the discrete variant of
diffusion on an arbitrary lattice, a topic interesting in its own
right. Diffusion on regular lattices is ubiquitous, and diffusion on
random graphs has (among other things) also been studied in the
context of glassy relaxation \cite{bray}. We are therefore inspired to
investigate what happens on the small-world model, which interpolates
between these two extremes. Previously a lot of interest has also been
seen in the related problem of diffusion on fractals (see for example
\cite{blumen1,blumen2,bunde,goyet} and references therein). As we
proceed to show, diffusion on Cayley trees \cite{kohler,cassi1,cassi2}
shows also features closely related to the present
problem. Furthermore, the motion of charge carriers or of excitons
over polymer chains, where steps between spatially close sites can
connect regions far apart along the chemical backbone, also involves
global shortcuts \cite{sokolov1,sokolov2}. 

The transition probabilities $W_{i,j}$ in Eq. (\ref{master})  are as
follows: First $W_{i,j}=0$ if there are no bonds between $i$ and
$j$. For $i$ connected to $j$ by one or more direct bonds, $W_{i,j}$
is proportional to the number of such bonds. The same holds for the
probability of remaining at the same site after one time unit, i.e. we
allow ``sticking''. Formally

\begin{equation}
\label{transprob}
W_{i,j}=\frac{z_{i,j}+\delta_{i,j}}{z_j+1}
\end{equation}

\noindent In this equation, $z_{i,j}$ is the number of bonds between
the two sites $i$ and $j$, and $z_i$ is the total number of bonds
emanating from vertex $i$, i.e. the coordination number of the
site. Hence $z_i=\sum_j z_{i,j}$. Note that the $z_{i,j}$-values are
determined both by the additional wiring as well as by the underlying
lattice. The $\delta_{i,j}$ and the $1$ in the denominator appear
because we allow for the possibility of the walker to remain at site
$i$ during a time step. This procedure renders the numerically
determined $P(i,n)$ smoother in $n$. We remark that the rates defined
according to Eq. (\ref{transprob}) are not symmetrical in $i$ and $j$,
i.e. in general $W_{i,j}\neq W_{j,i}$.

The algorithm we have used is the exact (cellular automaton)
enumeration of random walks \cite{bunde}, corresponding to the
implementation of Eq.  (\ref{master}). All the results plotted are
averaged over $200$ disorder configurations.  We have worked mostly
with the value $k=1$. This is also the value implied if we do not
state otherwise. 

\begin{figure}
\unitlength=1cm
\begin{center}
\begin{picture}(6,5.8)
\put(-2.2,6.6){
\includegraphics{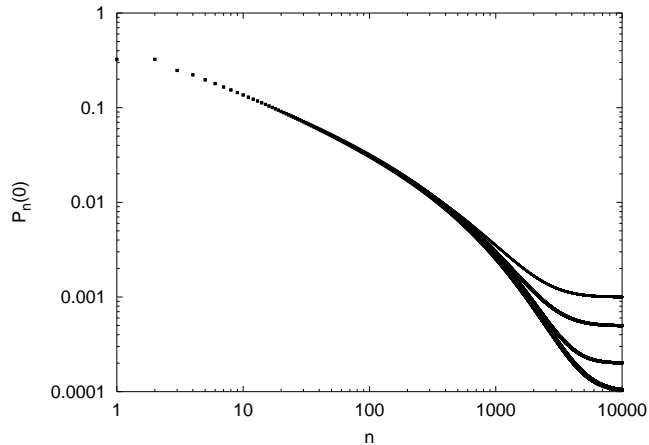}}
\end{picture}
\end{center}
\caption{The relaxation or probability of presence at the origin
  $P_n(0)$ as a function of number of steps for $p=0.05$ and several
  system sizes $L$, which from upper to lower right are $L=1000$,
  $L=2000$, $L=5000$ and $L=10000$.  
  } 
\label{varyLp1}
\end{figure}

We focus on the probability $P(i,n|i,0)$ that a particle initially at
site $i$ is found at the same site just after the $n$th step. In the
figures below we plot $\langle P(i,n|i,0)\rangle$, i.e. the average of
$P(i,n|i,0)$ over the different realisations of the small-world
lattice. Since all sites are equivalent  in an ensemble of small-world
networks, this quantity does not depend the particular site $i$
chosen, and we hereafter denote it by $P_n(0)$.  In Fig. \ref{varyLp1}
we have chosen $p=0.05$ and plotted $P_n(0)$ in double-logarithmic
scales for system sizes ranging from $L=1000$ to $L=10000$. This
allows us to examine the dependence of $P_n(0)$ on the size of the
system. 

From Fig. \ref{varyLp1} we infer that initially all the curves fall on
one curve and that for large $n$ they saturate at their respective
equilibrium values, $1/L$. However, Eq. (\ref{transprob}) implies an
inhomogenous equilibrium distribution

\begin{equation}
\label{eq}
P^{eq}(i)\propto (z_i+1).
\end{equation}

\noindent Therefore $P(i,\infty|i,0)$ depends on the specific small-world
realisation, and will fluctuate from realisation to realisation around
its average value $1/L$.

\begin{figure}
\unitlength=1cm
\begin{center}
\begin{picture}(6,5.8)
\put(-2.2,6.6){
\includegraphics{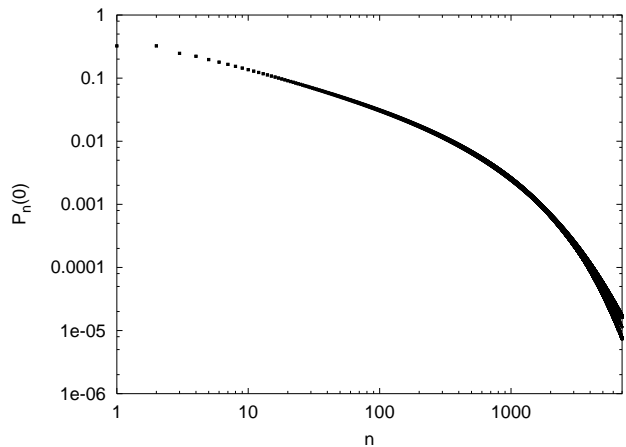}}
\end{picture}
\end{center}
\caption{Plot of $P_n(0)-P_\infty(0)$ as a function of $n$, the number of
  time steps for $p=0.05$ and $L$ as in Fig. \ref{varyLp1}. The curves
  fall nicely on a master curve.}
\label{pvaryLp1}
\end{figure}

\begin{figure}
\unitlength=1cm
\begin{center}
\begin{picture}(6,5.8)
\put(-2.2,6.6){
\includegraphics{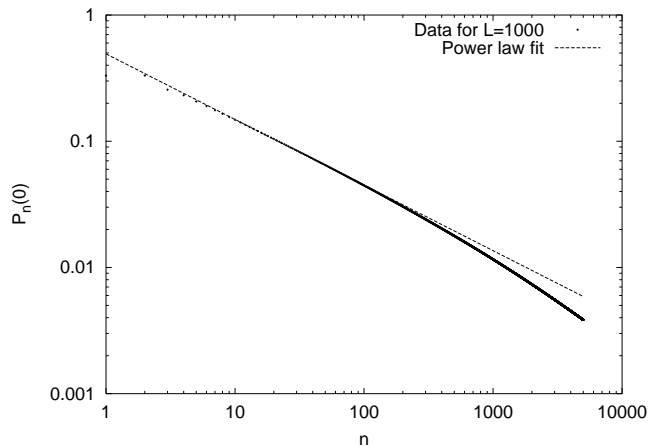}}
\end{picture}
\end{center}
\caption{$P_n(0)$ for $p=0.01$ as well as a power law
  approximation. The fit in the region $10<t<100$ gives as
  least-squares-fit exponent $-0.52$. }
\label{apppow}
\end{figure}

To find out how much of the behavior is due to finite size effects,
we subtract from each average curve in Fig. \ref{varyLp1} its
corresponding average equilibrium value
$P_\infty(0)\equiv 1/L$, and replot $P_n(0)-P_\infty(0)$ in
Fig. \ref{pvaryLp1}.  
\noindent From Fig. \ref{pvaryLp1} we see that all curves collapse nicely onto
what we view as representing $P_n(0)$ on small-world networks in the
limit $L\rightarrow\infty$. Both Figs. \ref{varyLp1} and
\ref{pvaryLp1} display initially a quasi-linear decay in the chosen
double logarithmic scales, and this may be viewed as being an
approximate power law decay. Depending on $p$, the
exponents range from around $-0.5$ for the smallest $p$ to around
$-0.6$ for the largest. This regime is followed by a steeper decay at
larger $n$. To highlight the power law character we have plotted in
Fig.\ref{apppow} $P_n(0)$ for $p=0.01$. As is evident from the figure,
the power-law domain extends well over two orders of magnitude
in $n$. 

\noindent The results can be understood qualitatively in the following way: For
a fractal one has \cite{alexander}

\begin{equation}
\label{fractal}
P_n(0)\sim n^{-d_s/2},
\end{equation}

\noindent where $d_s$ is the spectral dimension. Thus the initial
decay in Figs. \ref{varyLp1} to \ref{apppow} follows that of a fractal
with a $d_s$ close to $1$, i.e. that of a quasi $1$-d system.  This is
reasonable given our construction: for sufficiently small $p$ and
small $n$, only relatively few random walkers encounter any long-range
connections (shortcuts). Therefore in the beginning the behavior of
$P_n(0)$ closely reflects the character of the underlying $1$-d
lattice.  However for larger $n$, the random walkers probe larger and
larger portions of the graph, and thus follow more and more
shortcuts. This speeds up progressively the decay of $P_n(0)$ as more
regions at larger and larger length scales are visited, and the
fractal picture is lost. One would thus expect that the concept of a
$d_s$ begins to be invalid when the random walkers visit enough short
cuts, i.e. when the $1$-d diffusion extends longer than the typical
distance between shortcuts. This is the fundamental length scale $\xi$
of small-world networks, besides the lattice constant, which is less
important here. In our case we have
\begin{equation}
\label{lengthscale}
\xi=p^{-1},
\end{equation}
\noindent measuring $\xi$ in units of the lattice constant. For
diffusion on scales smaller than $\xi$ one furthermore has in terms of
the diffusion constant $D$ of the regular lattice $\xi^2\sim 2Dn$, so
that 

\begin{equation}
\label{timescale}
n\sim\frac{1}{2Dp^{2}}.
\end{equation}

\noindent Given that we allow random walkers to stay at a site during
a time step,  $D=1/3$ and thus $n=2/3p^{-2}$. However some walkers do
encounter shortcuts at length-scales below $\sim p^{-1}$, and
numerically the cross-over to a region that does not have approximate
power-law character is seen to take place earlier than $n\sim
p^{-2}$.

\begin{figure}
\unitlength=1cm
\begin{center}
\begin{picture}(6,5.8)
\put(-2.2,6.6){
\includegraphics{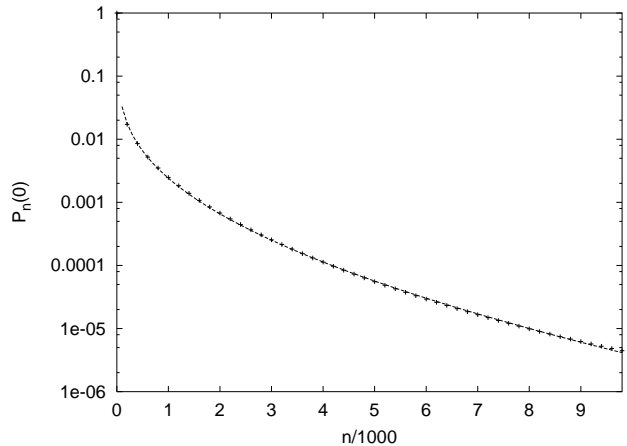}}
\end{picture}
\end{center}
\caption{$P_n(0)-P_\infty(0)$ for $p=0.05$ and $L=10000$ on a
    semi-logarithmic  scale. At longer times the decay appears to be
  slower than exponential. Also shown is a fit to a stretched
  exponential, indistinguishable from the data.}
\label{semilogp1}
\end{figure}

\noindent We turn now to the analysis of this region. To be able to
follow the it more closely, we  replot the results of
Fig. \ref{pvaryLp1} for $L=10000$ on semi-logarithmic scales in
Fig. \ref{semilogp1}. Evidently, the decay for larger $n$ is slower
than exponential. 

\noindent The decay of $P_n(0)$ is hence quicker than a power-law, but
slower than the one for Cayley-trees, for which one has (for
coordination numbers greater than $2$) \cite{cassi1,cassi2} 

\begin{equation}
\label{cayley}
P_n(0)\sim n^{-3/2} \exp\left[-Cn\right],
\end{equation}

\noindent where $C$ is a constant. Comparing this behavior to the one
displayed in Figs. \ref{varyLp1} and \ref{semilogp1} we remark that in
our case, for a relatively small number of steps, the decay goes
approximately as $n^{-\alpha}$ with $\alpha\gtrsim 1/2$, whereas at
larger $n$, a more adequate description would be a stretched
exponential, $\exp[-Cn^{\beta}]$. One may even suspect that the decay
in Fig. \ref{semilogp1} obeys $P_n(0)\sim
n^{-\alpha}\exp[-Cn^{\beta}]$.  A fit of the data in
Fig. \ref{semilogp1} to this functional form (keeping $\alpha \equiv
0.5$ fixed) is also shown; the fit turns out to be indistinguishable
from the numerical data, when we choose $\beta=0.56$ and $C=0.04$, as 
plotted in Fig. \ref{semilogp1}.  

\begin{figure}
\unitlength=1cm
\begin{center}
\begin{picture}(6,5.8)
\put(-2.2,6.6){
\includegraphics{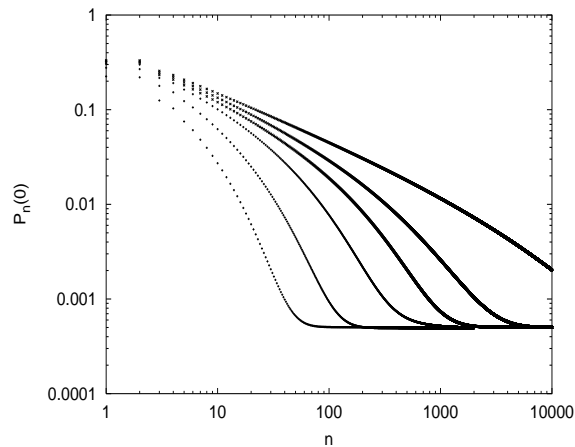}}
\end{picture}
\caption{The relaxation or probability of presence at the
    origin as a function of the number of steps for $L=2000$ and from
    upper right to lower left $p=0.01$, $0.05$, $0.1$, $0.2$, $0.4$
    and $0.8$.}
\label{varyp}
\end{center}
\end{figure}

\noindent We now consider the dependence of the decay on the value of
$p$. For this we plot in Fig. \ref{varyp} the decay law $P_n(0)$ for
$L=2000$ and $p$ ranging from $p=0.01$ to $p=0.8$. 
\noindent We note that the initial power-law like region
diminishes with increasing $p$. Furthermore, the plateau-region
$P_n(0)\simeq 1/L$ is reached earlier for larger $p$. This is in
accordance with our argument above, that the long-range connections
(short-cuts) interrupt the simple diffusion on the underlying lattice,
such that the crossover length decreases with increasing $p$
(c.f. also Eq. (\ref{timescale})).  As $p$ becomes large enough, the
power law regime practically disappears. This is so because the random
walker rapidly meets a short-cut. As before, the influence of finite
size effects can be reduced by plotting, as in Fig. \ref{pvaryLp1},
$P_n(0)-P_\infty(0)$. 

\begin{figure}
\unitlength=1cm
\begin{center}
\begin{picture}(6,5.8)
\put(-2.2,6.6){
\includegraphics{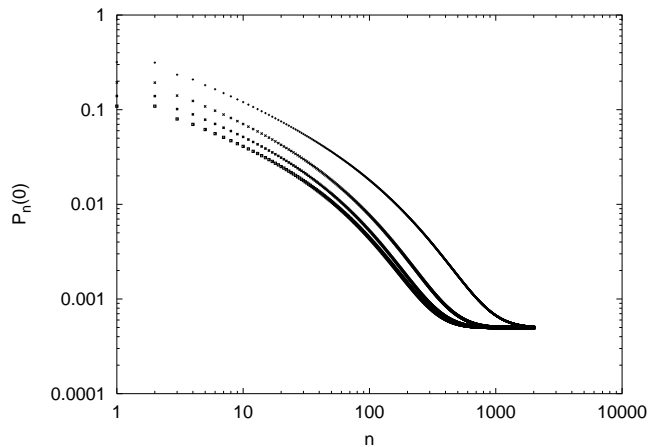}}
\end{picture}
\end{center}
\caption{$P_n(0)$ plotted for $L=2000$, $p=0.1$ and $k=1,2,3$ and $4$,
  from right to left.} 
\label{varyk}
\end{figure}

We have also performed simulations of the random walk on small-world
networks where the underlying lattice has a $k$ value larger than $1$.
In Fig. \ref{varyk} we plot the results for  $p=0.1$ and $L=2000$ in
the cases of $k=1$, $k=2$, $k=3$ and $k=4$.  The findings reproduce
the general picture: $P_n(0)$ behaves like a power law for small $n$,
while decaying more rapidly as $n$ gets larger. The curves for
different $k$ are mainly shifted with respect to each other, and the
network with the largest coordination number (largest $k$) also
displays the quickest relaxation. To be noted, however, is that the
case $k=1$ has the largest dynamical range and thus shows best the
decay forms, while also being the one simplest to implement; hence
$k=1$ may be the ideal small-world model. 

\section{Conclusions}
In this work we have studied numerically the behavior of random walks
on small-world lattices. Our work has focused on the  probability of
being at the initial site $P_n(0)$ as a function of the number of
steps $n$. This quantity is found to show a complex, very interesting
pattern: initially $P_n(0)$ displays a power-law, ``quasi-fractal''
regime. At larger $n$ a quicker decay takes over, reminiscent of
stretched exponentials. In this respect the $P_n(0)$ decay is
intermediate between the decays found for fractal structures and the
ones found for tree-like (loop-less) structures, exemplified here by
Cayley trees.

\renewcommand{\abstractname}{Acknowledgements}
\begin{abstract} 
The support of the DFG, of GIF grant I0423, and of the Fonds
  der Chemischen Industrie is gratefully acknowledged.
\end{abstract}


\begin{thebibliography}{99}


\bibitem{watts} D. J. Watts and S. H. Strogatz, {\it Nature} {\bf 393}
440 (1998).

\bibitem{amaral} L. A. Nunes Amaral, A. Scala, M. Barth\'el\'emy, and
  H. E. Stanley, {\it cond-mat/0001458}.

\bibitem{milgram} S. Milgram, {\it Psychol. Today} {\bf 2}, 60-67
(1967).

\bibitem{bollobas} B. Bollob\'as, {\it Random Graphs}, Academic
Press, London (1985).

\bibitem{barrat} A. Barrat and M. Weigt, {\it Eur. Phys. J. B} {\bf
13}, 547 (2000).

\bibitem{monasson} R. Monasson, {\it Eur. Phys. J. B} {\bf 12},
555 (2000).

\bibitem{newman1} M. E. J. Newman and D. J. Watts, {\it Phys. Rev. E}
{\bf 60} 6, 7332 (1999).


\bibitem{moukarzel2} C. F. Moukarzel, {\it Phys. Rev. E.}  {\bf
60} 6, R6263 (1999).

\bibitem{lago} L. F. Lago-Fern\'andez, R. Huerta, F. Corbacho and J. A. Sig\"uenza, {\it Phys. Rev. Lett.} {\bf 84} 20, 2758 (2000). 

\bibitem{newman2} M. E. J. Newman and D. J.Watts, {\it Phys. Lett. A}
  {\bf 263}, 341-346 (1999). 

\bibitem{moore} C. Moore and M. E. J. Newman, {\it
cond-mat/0001393}.

\bibitem{moukarzel1} C. F. Moukarzel and M. A. de
Menezes, {\it cond-mat/9905131}.

\bibitem{barthelemy} M. Barth\'el\'emy and L. A. Nunes
Amaral, {\it Phys. Rev. Lett.} {\bf 82} 15, 3180 (1999).

\bibitem{pandit} S. A. Pandit and R. E. Amritkar, {\it Phys. Rev. E}
{\bf 60} 2, R1119 (1999).

\bibitem{bray} A. J. Bray and G. J. Rodgers {\it Phys. Rev. B.} {\bf
    38} 16, 11461 (1988).

\bibitem{blumen1} A. Blumen and G. H. K\"ohler {\it
    Proc. R. Soc. Lond.} A {\bf 423}, 189 (1989).

\bibitem{blumen2} A. Blumen, J. Klafter, B. S. White and G. Zumofen
  {\it Phys. Rev. Lett.} {\bf 53} 14, 1301 (1984).

\bibitem{bunde} A. Bunde, J. Dr\"ager and M. Porto in {\it
    Computational Physics} ed. by K. H. Hoffmann and M. Schreiber,
    Springer (1996).

\bibitem{goyet} J-F. Goyet {\it Physics and Fractal Structures}, Springer (1996).

\bibitem{kohler} G. H. K\"ohler and A. Blumen {\it J. Phys. A} {\bf
23}, 5611 (1990).

\bibitem{cassi1} D. Cassi, {\it Phys. Rev. B} {\bf 45} 1, 454
(1992).

\bibitem{cassi2} D. Cassi, {\it Europhys. Lett.} {\bf 9}, 627
(1989).

\bibitem{sokolov1} I. M. Sokolov, J. Mai and A. Blumen {\it
Phys. Rev. Lett.} {\bf 79} 5, 857 (1997).

\bibitem{sokolov2} I. M. Sokolov, J. Mai and A. Blumen {\it
Czech. J. Phys.} {\bf 48}, 487 (1998).

\bibitem{alexander} S. Alexander and R. Orbach, {\it J. Phys. Lett.}
  {\bf 43}, L625 (1982).

\end{thebibliography}
\end{document}